\documentclass[reprint,amsmath,amssymb]{revtex4-1}

\usepackage{graphicx}
\usepackage{dcolumn}
\usepackage{bm}
\usepackage{color}
\usepackage[colorlinks=true,
            linkcolor=black,
            urlcolor=black,
            citecolor=blue, breaklinks=true]{hyperref}

\DeclareMathOperator{\sech}{sech}
\DeclareMathOperator{\csch}{csch}

\begin{document}

\title{Observational assessment of the viability of de Sitter G\"odel de Sitter phase transition}

\author{Sh. Khodabakhshi}
\email{sh\_khodabakhshi@ut.ac.ir}
\affiliation{
Department of Physics, University of Tehran, Tehran, Iran
}%

\author{M. Farhang}
\email{m\_farhang@sbu.ac.ir}
\affiliation{
Department of Physics, Shahid Beheshti University, Velenjak, Tehran 19839, Iran
}%
 
\author{A. Shojai}
\email{ashojai@ut.ac.ir  (corresponding author)}
\affiliation{
\hspace{0.1cm} Department of Physics, University of Tehran, Tehran, Iran
}%

\author{M. S. Esmaeilian}
\email{mu.esmaeilian@mail.sbu.ac.ir}
\affiliation{
\hspace{0.1cm} Department of Physics, Shahid Beheshti University, Velenjak, Tehran 19839, Iran
}%

\author{R. Moti}
\email{r.moti@ut.ac.ir}
\affiliation{
\hspace{0.1cm} Department of Physics, Faculty of Science, Ferdowsi University of Mashhad, P.O. Box 1436, Mashhad, Iran
}%

\date{\today}

\begin{abstract}
de Sitter--G\"odel--de Sitter phase transition(dGd) is a possible geometrical phase transition in the very early universe. It induces fluctuations with possibly observable traces on matter and radiation fields. Here we present a simulation based on dGd to investigate possible perturbations which could be along with the standard inflationary fluctuations in the cosmic microwave background (CMB) and distribution of the large-scale structure. The power spectrum of perturbations is characterized by a parameter pair, labeled here as  ($p_1, p_2$). 
With  {\it Planck} observations we find $p_1=0.008^{+0.003}_{-0.008}$ and $p_2= 0.002^{+0.001}_{-0.002}$ consistent with pure inflationary power spectrum and no hint for the dGd transition. Also, it is estimated future large-scale surveys such as Euclid and SKA can further tighten the constraints up to an order of magnitude and probe the physics of the early universe with much higher precision.
\end{abstract}

\maketitle

\section{Introduction}

Rotation is a universal phenomenon observed on a wide variety of scales in high-energy physics and astrophysics. Also, the origin of rotation is one of the most intriguing issues of cosmology. 
 Thanks to many redshift large area surveys\citep{davis1982survey, colless20012df, york2000sloan}, our knowledge about structure formation have increased, and now we know that most structures in cosmology are in a rotating state. 

Any theory of structure formation should explain the presence of rotation and clarify its origin\citep{korotky2020quest}. The later growth of structures in the nonlinear regime could cause structures to spin without any initial angular momentum. In the linear regime, however, a mechanism is required to initiate the rotation. Structures on scales of several Mega Parsecs like filaments, walls and voids provide an environment to test theories of linear and quasi-linear regimes\citep{bernardeau1996large, conroy2005deep2, croton2008red, sousbie2007three, bond2010crawling, choi2010tracing}. Surprisingly a recent observational effort gives some evidence for the spin of cosmic filaments\citep{Wang:2021axr}.

Theoretically, we know rotation is usually associated with vorticity. Therefore it can not be generated in a perfect fluid in perturbation theory and inflation scenarios. In particular, as the universe expands, any primordial vorticity will be redshifted away and would not be affected by the density perturbation growth\citep{Lu:2008ju}. Tidal Torque Theory (TTT) is the traditional formalism for analyzing the initial spin in the framework of the standard model of cosmology\citep{hoyle1951origin, Peebles:1969jm, White:1984uf}. Most literature follow TTT to explain the origin of rotation in linear and quasi-linear regimes, but the main issue is still an open question\citep{porciani2002testing, van2016zeldovich, pichon2014galactic, kim2022unexpected}.

Another suggestion for the origin of rotation backs to G. Gamow\citep{gamow1946rotating}. It is possible to consider the universe to be born with spin. Some cosmological models assume that the
primordial universe not only expands but also rotates\citep{Godel:1949ga, barrow1985universal, sivaram2012primordial}. But it has to be noted that due to the observational assessment, a global rotation is less than $ 10^{-13} rad \ y^{-1} $\citep{li1998effect}, and thus the idea of global rotation is not acceptable.

On the other hand, the early universe is a unique laboratory to test theories of high energy physics which are inaccessible to earth-bound experiments. Among these theories are possible cosmological phase transitions at different epochs depending on the energy scales involved (e.g. see \citep{Gangui:2001wc, Hindmarsh:1994re, Vilenkin:2000jqa}). If these transitions leave observable cosmological imprints, e.g. if they generate fluctuations on the CMB  radiation and matter fields, they would have the chance to be tested against data while their parameter $\beta$ would be constrained (see Ref. \citep{Ade:2013xla} for constraints on the cosmic string tension from Planck CMB observations).

Among plausible phase transitions in the early universe is de Sitter--G\"odel--de Sitter (dGd) phase transition which is a quantum phase transition of spacetime \citep{Khodabakhshi:2015ost}.
The dGd scenario assumes a scalar field in a de Sitter background would experience a phase transition to a rotating G\"odel geometry and slowly rolls back to the de Sitter phase.  

The dGd transition could be a possible source of initial rotation for large structures. This is particularly of interest since as mentioned before simple inflationary theories do not seed vector modes and therefore no initial spin is expected for the largest scale structure in pure inflationary cosmologies. As well, although TTT is the standard method, it is not the ultimate answer to the origin of rotation and is dependent on lots of numerical simulations and methods.

Quantum field theory calculations at finite temperature show that dGd second-order phase transition has a chance to occur at high temperatures. Then the transition probability depends on the rotation parameter of the G\"odel phase ($\alpha$), increasing as $\alpha$ decreases. 
 This rotation will be induced on the trajectories
 of test particles. Simulations show local congruence of particles has nonzero induced rotation while the average global rotation is almost zero\citep{Khodabakhshi:2015ost}.  

Furthermore, it was shown that Casimir force in the dGd transition induces inhomogeneities in the matter and radiation fields, possibly observable in the CMB radiation or large-scale structure data \citep{Khodabakhshi:2017bqx}. The predictions of the dGd transition can therefore be directly tested against the existing data. Observational assessment of the viability of this theory and estimating the model parameters are the main goals of this paper.

This paper is structured as follows:  We explain and review the dGd model in sections 2 and 3. In Section 4, we
simulate primordial seeds of inhomogeneities produced
by the dGd transition and assess the trace of primordial seeds (alongside inflationary perturbations)
on CMB anisotropies. Also, we explore how future large-scale surveys improve the bounds of model parameters. Finally, in Section 5, we discuss our conclusion.

\section{Review of \lowercase{d}G\lowercase{d} phase transition (Global features)} 

The extremely high temperature of the early universe allows for a scenario where a quantum mechanical phase transition could change the geometry of spacetime, which could occur due to a phase transition in the potential of a scalar field. Since 
 \begin{equation}
p_{\phi}=\frac{1}{2}\dot{\phi}^{2}-V(\phi)-\frac{1}{2}\left|\vec{\nabla}\phi\right|^{2}
\end{equation} 
\begin{equation}
\rho_{\phi }=\frac{1}{2}\dot{\phi}^{2}+V(\phi)+\frac{1}{2}\left|\vec{\nabla}\phi\right|^{2},
\end{equation} 
a spatially constant and time independent scalar field $ \phi $, ($ \dot{\phi} =0$ and $ \vec{\nabla}\phi=0 $) , acts as a perfect fluid with energy density and pressure given by $ p_{\phi}=-\rho _{\phi} $ which is the equation of state of cosmological constant.
Therefore we can consider $ V(\phi) $ as an effective cosmological constant\citep {Hobson:2006se}. Thus, the effective potential in curved spacetime would be an apt tool to continue. The action of a scalar field in a curved spacetime $ g_{\mu\nu} $ is \citep{Mukhanov:2007zz}
\begin{equation}
S[\phi,g_{\mu \nu }]=\int d^{4}x\sqrt{-g}\left(\frac{1}{2}\partial_{\mu}\phi\partial^{\mu}\phi-V(\phi)\right) \ .
\end{equation}
Straightforwardly we can calculate the one-loop effective potential from
\begin{equation}
V^{(1)}_{\textrm{eff}}=-\frac{\Gamma^{(1)}}{\cal{V}}
\end{equation}
where $\cal{V}$ is the spatial volume. $ \Gamma^{(1)}=-\frac{i}{2}\ln(\mu ^{-2}\det G)$ is the one-loop effective action, $G=\square +V''$ and  $ \mu  $ is introduced for dimensional considerations. Also $ \square $ operator is defined as $  \square \phi= |g|^{-1/2}\partial_\mu(|g|^{1/2}g^{\mu\nu}\partial_\nu\phi) $ for a surrounding metric $ g_{\mu\nu} $.

Now we briefly review the model. The dGd requires a situation where the background geometry of the universe would change from de Sitter to G\"odel for some small enough time and then quickly to de Sitter through a thermal phase transition of the scalar field. 

The dGd phase transition is supposed to occur just after inflation. The scalar field could be the inflaton field. Since we are working at a finite temperature regime and the wick rotation in curved space can only be adopted for stationary spacetimes, the model is restricted to the static chart of de Sitter, so that wick rotation could be applied. The static patch of de Sitter spacetime is described in $(t,\chi,\theta,\xi)$ coordinates by \citep{Fursaev:1993hm}
\begin{equation}
  \rm  ds^2=\cos^2\chi d\tau^2+a^2(d\chi^2+\sin^2\chi d\theta^2+\sin^2\chi\sin^2\theta d\xi^2)
\end{equation}
where  $a^2=2/\Lambda$ is the spatial scale and $ -\infty<t=-i\tau<+\infty $, $ -\pi<\chi<\pi $, $ 0<\theta$, $\xi<\pi $. The range of periodic parameter $ \tau $ is from $ 0 $ to $ \beta $. We assume the classical potential for $\phi$ is
 \begin{equation}
 V=\frac{1}{2}\sigma^2\phi^2+\frac{1}{24}\lambda\phi^4
\label{classical v}
 \end{equation}
With $\sigma$ and $\lambda$ to be an arbitrary mass scale and a dimensionless coupling constant, respectively. Using the zeta-function $(\zeta)$ regularization method, the corresponding effective potential to (\ref{classical v}) is
\begin{multline}
 V_{\textrm{eff}}(\phi ,\beta )=V(\phi)- \\ 
 \frac{1}{2\beta \cal V}\left[\zeta'(0,\beta)+\log(\mu^{2}a^{2})\zeta(0,\beta)+\log(V''(\phi)\mu^{-2})\right]
 \end{multline} 
where $\beta=(k_BT)^{-1} $ \citep{Elizalde:1994gf}. Following calculations of \citep{Khodabakhshi:2015ost} we find the effective potential in de Sitter background
\begin{multline}
\frac{\lambda}{2\sigma^4}V_{\textrm{eff}}=\frac{\lambda}{2\sigma^4}V_0+\frac{1}{2}x^2+\frac{1}{12}x^4 \\
-\gamma T\bigg\{0.62\Delta-0.22\Delta^2+0.075 T+\frac{T^3}{120}\\
+\frac{1}{T}\left(-0.07+2.7\Delta^2-0.15\Delta\right)\\
 +\left(0.22-\frac{2.7}{T}\right)\left(\frac{51-60T^2-8T^4}{240}-\frac{2T^2-3}{2}\Delta+\Delta^2\right)\\
+\log[\mathfrak{a}(1+x^2)]\bigg\} \
 \label{desitterpotential}
\end{multline}
where $\mathfrak{a}=a^2\sigma^2$, $\gamma=\frac{3\lambda}{32\pi^2\mathfrak{a}^2}$, $x=\sqrt{\frac{\lambda}{2}}\frac{\phi}{\sigma}$, $\Delta=\frac{9}{4}-\mathfrak{a}(1+x^2)$, and $T=\frac{\beta _{H}}{\beta}=\frac{2\pi a}{\beta}$. 

\begin{figure}
  \begin{center}
    \includegraphics[height=5cm]{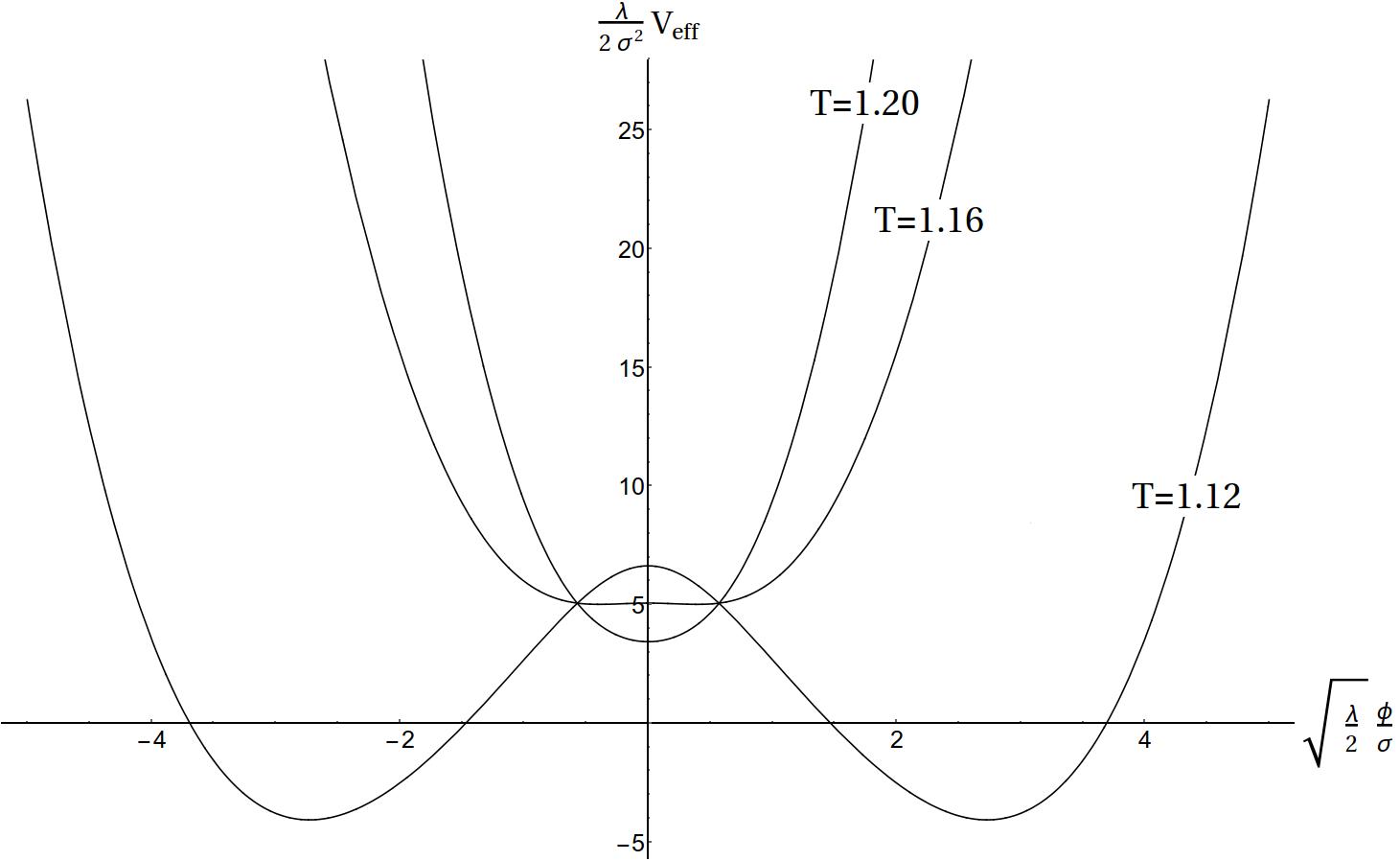}
    \caption{The one-loop potential for different values of temperature, $\mathfrak{a}=2$ and $\gamma=10$ taken from \protect \citep{Khodabakhshi:2015ost}.}
       \label{dgd0}
  \end{center}
\end{figure}

\begin{figure}
  \begin{center}
    \includegraphics[height=5cm]{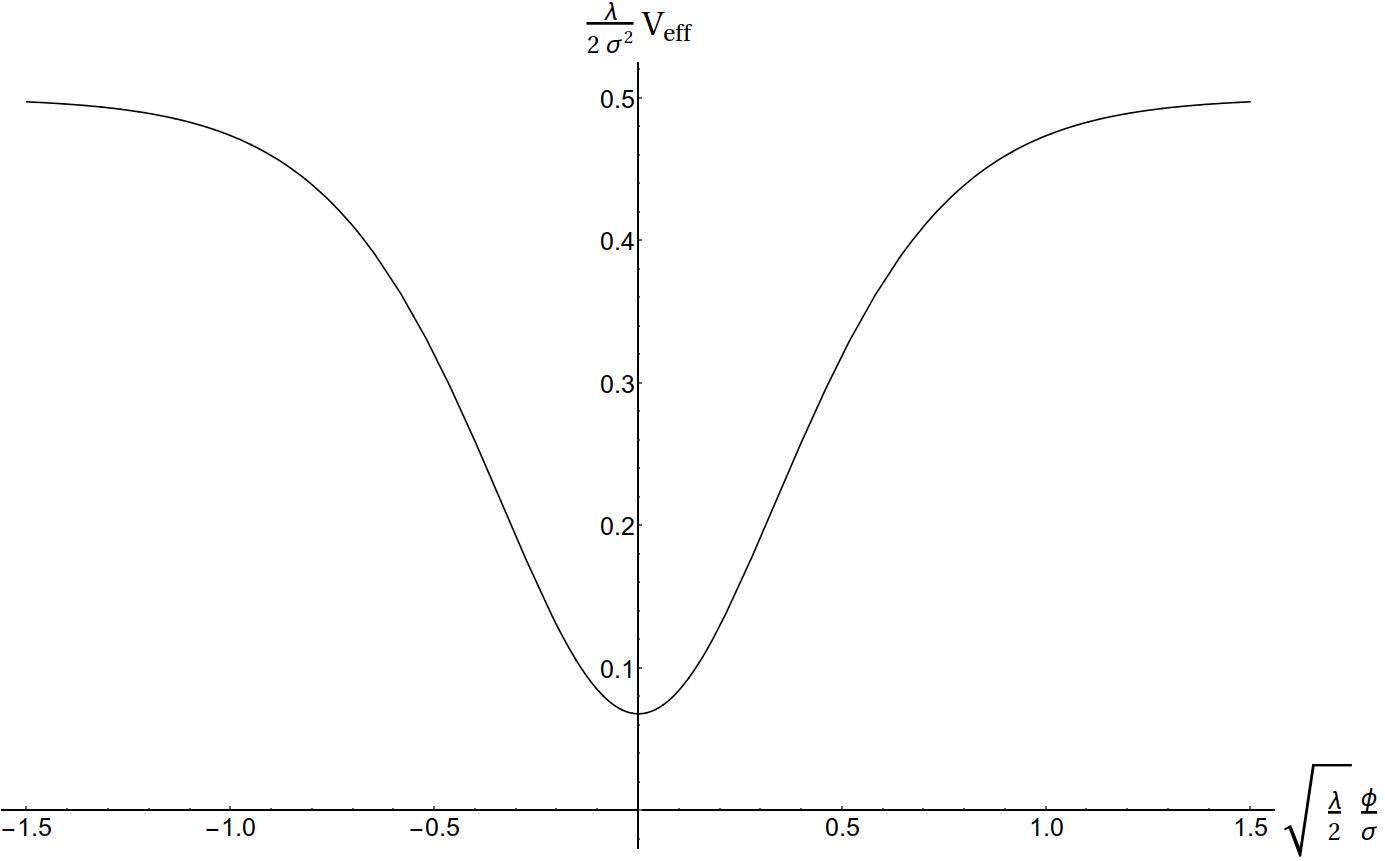}
    \caption{one-loop effective potential in G\"odel spacetime with $\bar\alpha=3$, $\bar\gamma\bar T=10$, and $\frac{\lambda V_0}{2\sigma^2}=0.5$, taken from \protect \citep{Khodabakhshi:2015ost} .}
       \label{godelpotential}
  \end{center}
\end{figure}

This potential has extremum at
\begin{equation}
x=0
\end{equation}
and at the roots of the equation
\begin{multline}
  1+\frac{1}{3}x^3-\gamma T\bigg\{ -1.24\mathfrak{a}+0.44\Delta-\frac{10.8}{T}\Delta+\frac{0.3\mathfrak{a}}{T}\\
  \mathfrak{a}\left(0.22-\frac{2.7}{T}\right)(2T^2-3-4\Delta)+\frac{2}{1+x^2}\bigg\} \ .
\end{multline}
It is clear that if this equation has no real root, the potential is a harmonic-type. Otherwise, it has a Mexican hat shape. 
There are some regions of parameters $ \mathfrak{a} $, $ \gamma $ and temperature $ T $ which allow a Mexican hat shape for Eq.~\ref{desitterpotential}. Typical shape of the effective potential is plotted in figure~\ref{dgd0} for some different values of temperature, $\mathfrak{a}=2$ and $\gamma=10$.
Therefore, during the universe cooling, we can find a critical temperature in which the scalar field experiences a phase transition so that the effective potential is negative. At this temperature and when the depth of the Mexican hat is equal to $-\Lambda=\rho_{\textrm{dust}}/2$, we assume the universe has G\"odel geometry. Then we can evaluate the one-loop potential with G\"odel  metric, which is an exact dust solution of Einstein field equations via a negative cosmological constant, described in Cartesian coordinates $(t,x,y,z)$ by \citep{Godel:1949ga}
\begin{equation}
 \rm  ds^2=(dt+\exp^{\sqrt{2}\alpha x/2}dy)^2-dx^2-\frac{1}{2}\exp^{\sqrt{2}\alpha x}dy^2-dz^2
  \label{godel}
\end{equation}
where $\alpha$ gives the angular momentum four-vector $\alpha ^{\beta}=(0,0,0,\alpha)$ of intrinsic rotation. Following calculations of \citep{Huang:1991eg}, we can calculate the effective potential in G\"odel spacetime
\begin{multline}
  V_{\textrm{eff}}=\bar V_0+ \frac{\sigma^4}{16\pi ^{3/2}}\left(\frac{\alpha^2}{8\sigma^2}+\frac{M^2}{\sigma^2}\right)^{3/2}\times\\
  \bigg\{\frac{2\sqrt{\pi}}{3}-\sum_{n,\ell }^{'}\left(2Z^{\nu-3/2}_{1}K_{3/2-\nu}(2Z_1)\right.\\
 \left.-Z^{\nu-3/2}_{2}K_{3/2-\nu}(2Z_2)\right)\bigg\}
\end{multline}
where $\bar V_0$ is a constant and $ M^2=\sigma^2+\frac{1}{2}\lambda\phi^2$. Prime on the summation means the term $ n=\ell=0 $ is neglected. $ K $ is the modified Bessel function, and $ Z $ is the Epstein zeta function  
\begin{equation}
Z_{1}=\pi\sqrt{\left(\frac{\alpha^2}{8}+M^{2}\right)\left(\frac{\ell^{2}}{4\pi^{2}\bar T^{2}\sigma^2}+\frac{8n^{2}}{\alpha^{2}}\right)}
\end{equation}
\begin{equation}
Z_{2}=\pi\sqrt{\left(\frac{\alpha^2}{8}+M^{2}\right)\left(\frac{\ell^{2}}{4\pi^{2}\bar T^{2}\sigma^2}+\frac{2n^{2}}{\alpha^{2}}\right)}
\end{equation}
where $\bar T^{-1}=\beta\sigma$. In $\bar\alpha=8\sigma^2/\alpha^2\gg 1 $ and $ \bar T\gg 1$ limits, the renormalized effective potential with dimensionless quantities is
\begin{multline}
\frac{\lambda}{2\sigma^4}V_{\textrm{eff}}=\frac{\lambda}{2\sigma^4}\bar V_0-\bar\gamma \bar T \left(1+\bar\alpha(1+x^2)\right)^{1/4}
\\
\times \exp\left(-\pi\sqrt{1+\bar\alpha(1+x^2)}\right)
\end{multline}
where $\bar\gamma=\frac{\lambda\alpha^2}{16\pi^2\sqrt{2}\sigma^2}$. The one-loop effective potential shape with these limits, is presented in figure~\ref{godelpotential}.
Here $\alpha$ is a nonzero real constant which shows the rotation rate of dust around the $y$-axis. The G\"odel phase can naturally initiate the rotation since it is a rotating spacetime. In this phase, $V_{\textrm{eff}}=-\rho_{\textrm{dust}}/2$ and has the role of a negative cosmological constant. Finally, the shape of the effective potential in the G\"odel background allows for the rolling of the scalar field such that the universe goes back to the de Sitter phase with a positive sign of effective potential.
 \begin{figure}
	\includegraphics[width=\columnwidth]{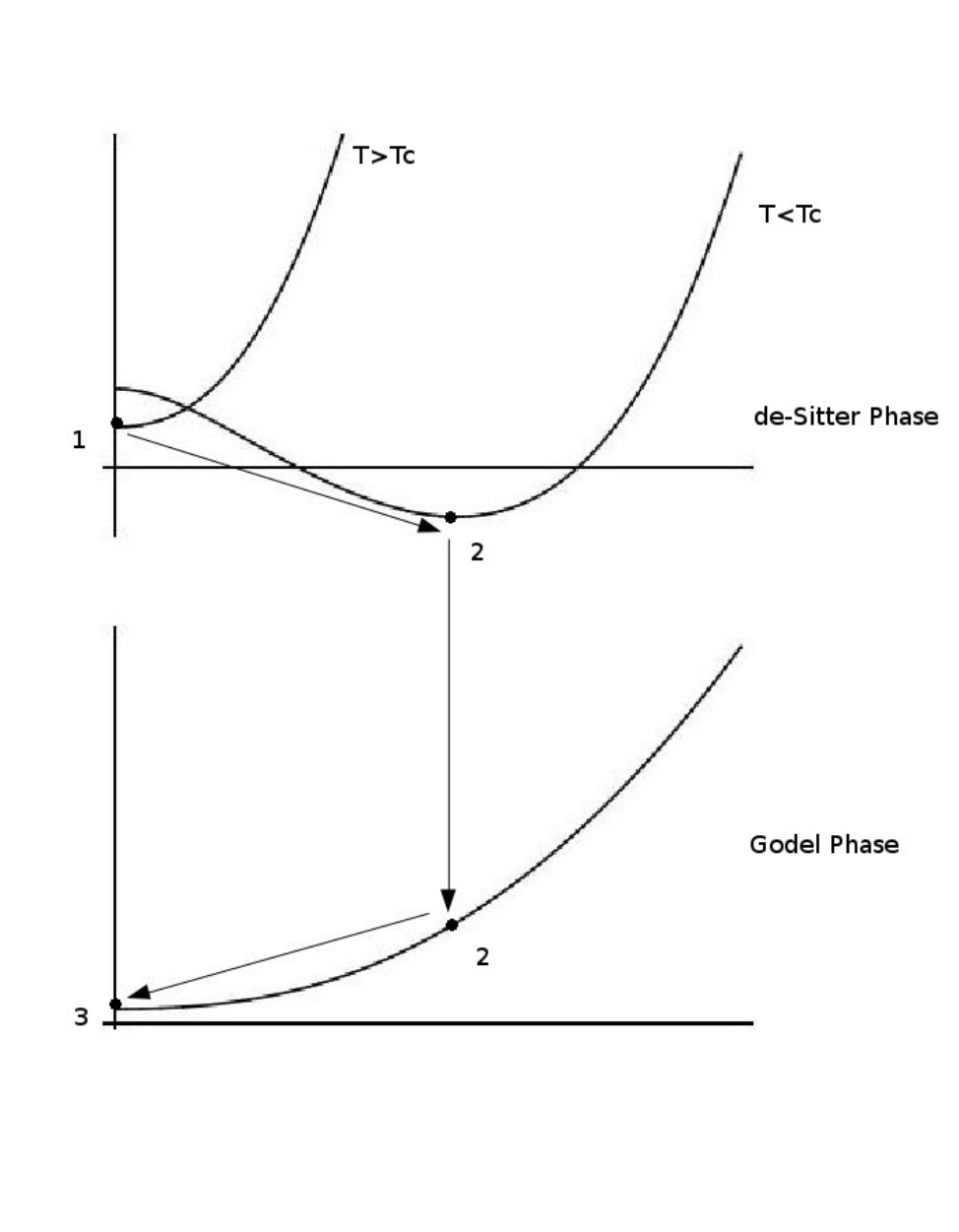}
    \caption{Typical scenario of the initial de Sitter, G\"odel and final de Sitter phases, taken from \protect \citep{Khodabakhshi:2015ost}.}
    \label{dgd1}
\end{figure}

Figure~(\ref{dgd1}) shows a typical scenario to understand the whole story in three steps. First, $V_{\textrm{eff}}$ is much larger than the dust density $V_{\textrm{eff}}\gg \rho_{\textrm{dust}}$ and is positive so it can act like a positive cosmological constant. After enough cooling due to the expansion, the universe can reach a critical temperature $\beta_c$ where $\left. \frac{d^2V_{\textrm{eff}}}{d\phi^2}\right|_{\phi=0}=0$. Below $ \beta_c $, the scalar field can experience a phase transition to the Mexican hat potential with a negative cosmological constant which describes a G\"odel phase. Then the scalar field can roll down the potential until $\phi=0$. After  $\tilde t\simeq\frac{\phi}{\dot\phi}\simeq\frac{\sigma}{\sqrt{\lambda\Lambda}}$ (which is the time duration of dGd phase transition), the universe would go back to a de Sitter phase again.

The impact of dGd on the equations of motion of a test particle is explored in \citep{Khodabakhshi:2015ost}. As intuitively expected, a  particle that enters the first de Sitter phase with a non-rotating trajectory exits to the final de Sitter phase while it is rotating. In other words, this phase transition induces rotation in the motion of test particles. The value of the induced rotation depends on the particle position in the de Sitter space, its distance from the symmetry axis of G\"odel, and the initial velocity, and it is of order $ \sqrt{\Lambda} $. 
This mechanism also works for the congruence of particles. Simulations show a local congruence of particles would obtain nonzero local induced rotation. Also if we divide the universe into \textit{cells} and simulate the dGd transition based on the quantum tunneling probability and the randomness of the symmetry axis of G\"odel space direction, we will find the average global induced rotation is nearly zero, as expected. See Figures 5 and 7 of \citep{Khodabakhshi:2015ost}.

\section{Local features of \lowercase{d}G\lowercase{d}}

\subsection{Israel junction condition}

\begin{figure}
   \begin{center}
  \includegraphics[width=\columnwidth]{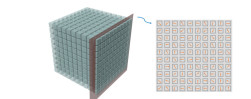}
     \caption{universe as a 3D lattice (left). The cells experience rotations around  random axes (right).(taken from \protect \citep{Khodabakhshi:2017bqx})}
        \label{fig:lattice}
   \end{center}
   \end{figure}

Casimir force comes from zero-point oscillations of quantized field between two boundaries \citep{Bordag:2009zz}. We can assume the universe is built from a 3D lattice with $n^3$ cubic cells with side $d$ as Figure~\ref{fig:lattice}. Each cell could experience the dGd phase transition. Thus, if the rotation direction of neighboring rotating cells is different (Figure~\ref{fig:lattice}), then the Israel density at the boundaries of each of them, leads to a Casimir force. Isreal density is nothing but the jump at the intrinsic curvature \citep{poisson2002advanced}

\begin{equation}
S_{ab} \propto [K_{ab}]-[K] h_{ab} \ ,
\end{equation}
where  $[K_{ab}] = K_{ab}^{(2)}-K_{ab}^{(1)}$ and $ K_{ab}^{(i)} \equiv n_{(\alpha;\beta)} e^{\alpha \ (i)}_{a} e^{\beta \ (i)}_{b}$  .

In our model, the jump in the intrinsic curvature is a consequence of the random direction of  G\"odel cells. To have a rare estimation of this density, we use the G\"odel metric in cylindrical coordinates,
\begin{multline}
d\rm s^2 = \frac{2}{\alpha^2} \bigl[ dt^2 - dr^2 -dz^2 +(\sinh^4r-\sinh^2r) d\phi^2 \\
\rm +2\sqrt{2} \sinh^2 r d\phi dt \bigr] \ .
\end{multline}
Without loss of generality, we assume a hypersurface between two cells is determined by the normal $n = (0,\sin\phi,\cos\phi,0)$. Thus, at one side the extrinsic curvature $K^{(1)}_{ab}$ is
\begin{widetext}
\begin{small}
\begin{equation}
\begin{pmatrix}
  0 & \sqrt{2}\cos{\phi}\csch{r}\sech{r} & -\sqrt{2}\sin{\phi}\cosh{r}\sinh{r} \\ 
  \sqrt{2}\cos{\phi}\csch{r}\sech{r} & 0 & \dfrac{1}{2}\cos{\phi}-\cos{\phi}\csch{r}\sech{r} \\
  -\sqrt{2}\sin{\phi}\cosh{r}\sinh{r} & \dfrac{1}{2}\cos{\phi}-\cos{\phi}\csch{r}\sech{r} & -\dfrac{1}{4} \bigl(4-4\sinh{2r}+\sinh{4r} \bigr) \sin{\phi}
\end{pmatrix}
\end{equation}
\end{small}
\end{widetext}
while on the other side this quantity is the one which is fixed by the rotated local tetrads concerning the previous one. Up to our simple estimation, it would be acceptable and more fanciable if we rotate the deviation of the normal vector, $n_{(\alpha;\beta)}$, rather than the tetrads. Thus, for $n' = (0,\cos(\phi),-\sin(\phi),0)$, the $K^{(2)}_{ab}$ is
\begin{widetext}
\begin{small}
\begin{equation}
\begin{pmatrix}
  0 & -\sqrt{2}\sin{\phi}\csch{r}\sech{r} & -\sqrt{2}\cos{\phi}\cosh{r}\sinh{r} \\ 
  -\sqrt{2}\sin{\phi}\csch{r}\sech{r} & 0 & \dfrac{1}{2}\bigl(\sin{\phi}\csch{r}\sech{r}+(\csch{r}\sech{r}-1)\sin{\phi} \bigr) \\
  -\sqrt{2}\cos{\phi}\cosh{r}\sinh{r}  & \dfrac{1}{2}\bigl(\sin{\phi}\csch{r}\sech{r}+(\csch{r}\sech{r}-1)\sin{\phi} \bigr) & -\dfrac{1}{4} \bigl(4-4\sinh{2r}+\sinh{4r} \bigr)\cos{\phi}
\end{pmatrix}
\end{equation}
\end{small}
\end{widetext}
Therefore, the Israel density for this choice is about
\begin{multline}
S_{00} \propto \dfrac{1}{4}  \bigl(4-4\sinh{2r}+\sinh{4r}\bigr) \cos{\phi} \csch^2{r} \sech^2{r} \\
-\dfrac{1}{4}  \bigl(4-4\sinh{2r}+\sinh{4r}\bigr) \csch^2{r}\sech^2{r}\sin{\phi} \ .
\label{israel}
\end{multline}
Although the precise value of this quantity is obtained by calculating the Root mean square over all possible orientations of the neighboring cells, the nonvanishing result of Eq.~\ref{israel} is enough motivation for the following discussions.

\subsection{Casimir effect}

As discussed, the nonvanishing expression of the Israel junction condition shows after the dGd phase transition, the distance between local layers of matter can shrink in a direction perpendicular to the random G\"odel rotation axis in an inhomogeneous way. This can cause a Casimir effect in the G\"odel phase. The Casimir energy of a scalar field in the G\"odel background at finite temperature is \citep{Khodabakhshi:2017bqx}
\begin{align}
\bar E_{\textrm{\tiny Casimir}}\left(\bar d,\bar\beta\right)&= \bar E_0\left(\bar d\right)+\bar\Delta_{\textrm{F.T.}}\left(\bar d,\bar\beta\right) \ .
\end{align}
The zero temperature term $\bar E_0\left(\bar d\right)$, and the finite temperature contribution $\bar\Delta_{\textrm{F.T.}}\left(\bar d,\bar\beta\right)$ are respectively given by 
\begin{align}
\bar E_0(\bar d) \simeq -\frac{\pi}{24\bar d}-\frac{1}{4\sqrt{2}\bar d}\exp \left \{-2\left (\frac{1}{2}+\bar m^2\right )\frac{\bar d^2}{\pi}\right \},
\end{align}
and 
\begin{align}
\bar\Delta_{\textrm{F.T.}}\left(\bar d,\bar\beta\right)&=\frac{1}{\bar\beta}\ln  \bigg[ \left \{ 1-\exp\left (-\frac{2\left (\frac{1}{2}+\bar m^2\right )\bar d^2}{\pi}\right )\right \} ^{\frac{1}{2}} \nonumber \\
 \prod_{n_1,n_2,n_3,n_4=1}^{\infty}& \left \{ 1-\exp \left (-\bar\beta\sqrt{\frac{1}{2}+\bar m^2+n_1^2+\frac{\pi^2n_2^2}{\bar d^2}} \right ) \right \}\nonumber\\
\times & \left \{ 1-\exp \left (-\bar\beta\sqrt{\frac{1}{2}+\bar m^2+\frac{\pi^2n_3^2}{\bar d^2}}\right ) \right \} \nonumber \\
\times & \left \{ 1-\exp \left (-2\bar d\sqrt{\frac{1}{2}+\bar m^2+n_4^2}\right )\right \}^{\frac{1}{2}}  \bigg] \ .
\end{align}
Here $ \bar{\beta}=\beta\alpha$, $\bar{d}=\alpha d$, $\bar{m}=m/ \alpha $ and  $\bar E=E/\alpha$ are dimensionless quantities in  natural unites,  $ \alpha $ is the rotation rate of the G\"odel metric (see Eq.~\ref{godel}), $d$ is the distance between plates and $n_1,n_2,n_3,n_4$ are introduced in \citep{Khodabakhshi:2017bqx}. 

\begin{figure}
	\includegraphics[width=\columnwidth]{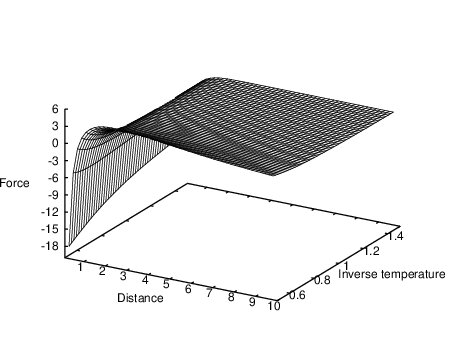}
    \caption{Normalized Casimir force, $ \bar{m}=1 $(taken from \protect \citep{Khodabakhshi:2017bqx}).}
    \label{casimirforce}
\end{figure}

Some direct calculation yields the expression for the Casimir force. The general behavior is plotted in Figure~\ref{casimirforce}.
 Here we only keep the major contribution to the force given by  $n_1=n_2=n_3=n_4=1 $. The  $ \bar{d} $-dependence of the Casimir force would then be  
\begin{equation}
\mathit{F}_{\textrm{\tiny{Casimir}}}\simeq \mathcal{O}(\frac{1}{\bar{\beta}^2\bar{d}^{1/2}})+\mathcal{O}(\frac{\bar{d}^{1/2}}{\bar{\beta}^2})+\mathcal{O}(\frac{1}{\bar{\beta}})+\mathcal{O}(\frac{1}{\bar{\beta}\bar{d}^2}), 
\end{equation}
or in terms of $ \alpha$:
\begin{equation}
\mathit{F}_{\textrm{\tiny{Casimir}}}\simeq\alpha^2\left\{ \mathcal{O}(\alpha^{-5/2})+\mathcal{O}(\alpha^{-3/2})+\mathcal{O}(\alpha^{-1})+\mathcal{O}(\alpha^{-3})\right\} \ .
\end{equation}
Asymptotically, in the limit of small rotations, we get
\begin{equation}
\mathit{F}_{\textrm{\tiny{Casimir}}}\simeq\frac{1}{\bar{\beta}\bar{d}^2} \ .
\label{force}
\end{equation}

The Casimir force is sensitive to the rotation angle of the G\"odel spacetime. We demonstrated that for two parallel plates with a separation comparable to the rotation of G\"odel spacetime ($\alpha$), the force becomes repulsive and then approaches zero.
This effect, when considered collectively due to many layers, could induce inhomogeneities that are potentially observable.  In the next section, we investigate the observable consequences of dGd transition produced by local Casimir forces.

It may be asked that despite the existence of closed time-like curves (CTCs) and causality violation properties of G\"odel spacetime, why such a  metric is chosen. Actually, to design a phase transition model to produce rotation we need a metric with intrinsic rotation and a non-vanishing cosmological constant. Meanwhile, we should choose a metric that contains dust matter and thus vacuum solutions of Einstein's equations like the Kerr metric\citep{teukolsky2015kerr} are not good for this purpose. Also, since the effective potential calculations and renormalization procedures are heavy enough, we restricted ourselves to the simplest rotating spacetime i.e. the G\"odel and not the G\"odel-type metrics\citep{rebouccas1983homogeneity} or Bianchi ones\citep{krasinski2001rotating}. 

One should not be worried about CTCs, because they are non-geodesics paths in G\"odel spacetime\citep{buser2013visualization, nolan2020causality} and are not an obstacle in the calculations. Furthermore there are some causal regions in G\"odel spacetime and we assumed the dGd phase transition takes place at those causal areas. However, it is important to mention that the main effect which was used in dGd phase transition is that at finite temperatures there are some critical points where the Casimir effect in G\"odel background becomes repulsive and this is argued in  Figures 1, 2, and 3 of \citep{santos2022thermal} and they have shown this behavior occurs in both causal and non-causal regions.

\section{Simulations and results}\label{sec:res}

 It was argued in the previous section that an early dGd phase transition (happening around the end of inflation)
 would generate the Casimir forces. These forces would generate potentially observable inhomogeneities in the universe.
 In this section, we first simulate the fluctuations in the inflaton filed by the dGd scenario and then consider them as possible seeds of inhomogeneities in the universe. Our goal is to assess their detectability in the observations of CMB anisotropies and large-scale structures.
 
Consider the universe as a 3D lattice with $n^3$ cubic cells (with side $d$) as schematically illustrated in Figure~\ref{fig:lattice}.
 We find $n=30$ to be a proper choice in this work, yielding converged results with reasonable computational cost.
 The location of each cell is represented by its center coordinates. 
 In the G\"odel phase each cell would experience some shrinkage, $\delta$, along a random direction.
 
 Generally, one can calculate $\delta$ using the geodesic equation of a test particle moving under the Casimir force. 
 However, since $\tilde{t}$ is considered to be small, the Newtonian approximation $ \delta \approx \frac{1}{2m} \mathit{F}_{\textrm{\tiny{Casimir}}}\tilde{t}^2$ would suffice. On the other hand, 
  using Eq.~\ref{force} yields 
   
    \begin{equation}
  \tilde{t}\simeq \frac{\sigma}{\sqrt{\lambda\Lambda}} \ ,
  \label{ttilde}
\end{equation}

implying the dependence of $\delta$ on these physically more informative quantities. The cosmological constant of the G\"odel phase, $\Lambda$,  depends on the G\"odel rotation parameter $\alpha$ through  $\Lambda=-\alpha^2 / 2$. 
Around the end of inflation (e.g., after about 60 e-foldings), $ \Lambda$ can be approximated by $ \Lambda =-4\pi \rho$. 
 Reasonable assumptions for the parameters of the theory give $ \frac{\delta}{d}\sim 10^{-8} $ \citep{Khodabakhshi:2017bqx}.

The rotation of cubes, therefore, generates fluctuations in the density field due to the reduction in the cell volumes from the Casimir effect. 
 We developed a Fortran code to simulate these dGd-induced inhomogeneities and generated $N_{\rm sim}=100$ realizations. Each cell experiences some rotation in a random direction and therefore suffers from Casimir-based shrinkage in its volume, estimated to be $\delta$.
 Also, adjacent cells would overlap in volume due to their random rotation, leading to changes in their densities. 
 To facilitate the computation of the volume overlaps, we divide each cell into $n_{\rm grid}^3$ (with $n_{\rm grid}=30$), and count the number of the fine cells sticking out of or coming into the boundaries of the original volume due to rotation.  
  The overall volume change of a cell, and therefore its density contrast against the background, is then calculated by taking into account both of these shrinkage and overlapping cell effects.  
 It should be noted that since the rotation is due to a quantum phase transition, more precise simulations should take into account the quantum tunneling nature of the transition and therefore the rotation. In this work, however, we ignored this effect for simplicity.
The result of each simulation would be an array of local density variations $ \Delta(\vec{x})=\delta\rho(\vec{x})/\rho(\vec{x}) $. One then gets the correlation function 
$\xi(\vec{r})=\langle \Delta(\vec{x}+\vec{r})\Delta(\vec{x})\rangle$ of the predicted primordial density field $ \Delta(\vec{x})$ 
where $\langle...\rangle$ represents averaging over the $N_{\rm sim}$ simulations.

The Fourier transform of the correlation function of the density field would give the power spectrum of the primordial field 
 $\langle\tilde{\Delta}(\vec{k})\tilde{\Delta}(\vec{k}') \rangle =(2\pi)^3\delta_{\text{D}}(\vec{k}+\vec{k}')\mathit{P_{\delta \phi}}(k)$, 
 where $\tilde{\Delta}(\vec{k})$ represents the Fourier transform of  $\Delta(\vec{x})$.  
 The required conversion from $P_{\delta \phi}$ (as directly calculated from simulations) to the primordial power spectrum ${\cal P}(k)$ of curvature perturbations is the same as  in standard inflationary scenarios, with the only difference being the shape of $P_{\delta \phi}$. 

\begin{figure}
   \begin{center}
    \includegraphics[width=\columnwidth]{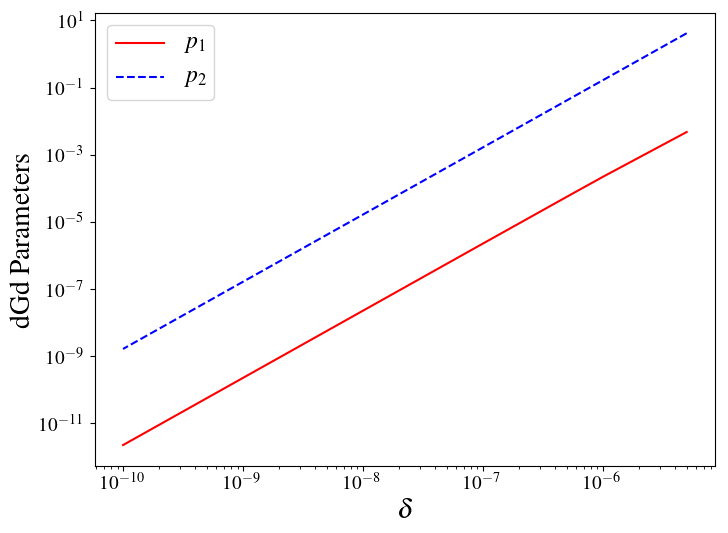}
     \caption{The dGd parameters $p_1$ and $p_2$ as  functions of $\delta$, the main free physical parameter of the scenario.}
        \label{pp}
   \end{center}
\end{figure}

By repeating the simulations for different values of $\delta$ which can be considered the main physical free parameter of the scenario, 
 we find that the dimensionless power spectrum for the induced curvature perturbations, ${\cal P}(k)$,  can be fitted by 

\begin{equation}
{\cal P}(k)=\big( p_0+p_1 (k/k_{\rm p})+\frac{p_2}{(k/k_{\rm p})^n}\big) \times 10^{-10} \ ,
\label{pk}
\end{equation}

where $k_{\rm p}=0.05 \rm{Mpc}^{-1}$ is the pivot scale for scalar perturbations and $p_{0,1,2}$ are dimensionless coefficients. 
 It turns out that the functional form of the fitted curve is quite insensitive to the choice of $\delta$ and $\delta$ only affects the parameter values. We also find that $n\approx 1$.
 Figure~\ref{pp} illustrates the dependence of the dGd parameters $p_1$ and $p_2$ on $\delta$ over a wide span. 

\begin{figure}
   \begin{center}
     \includegraphics[scale=0.43]{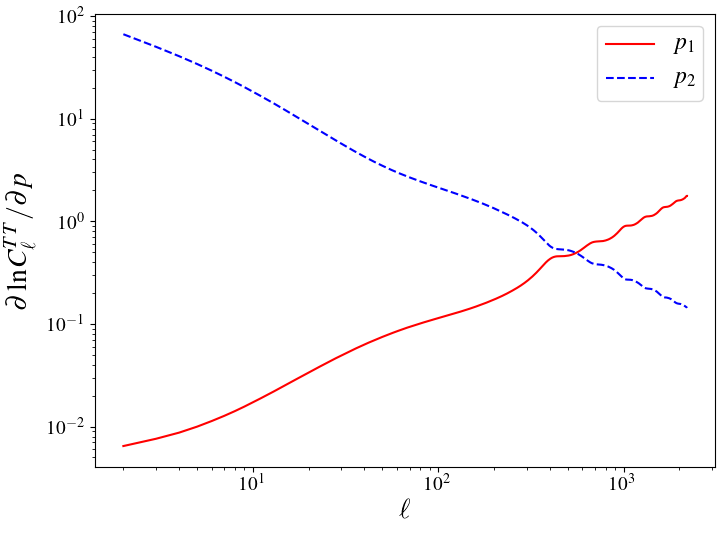}
     \includegraphics[scale=0.5]{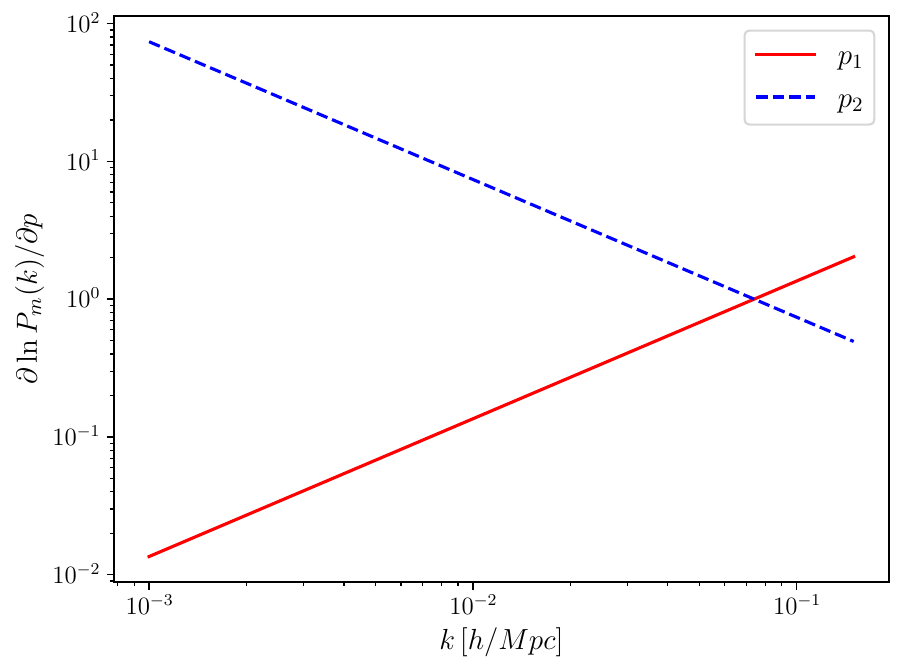}
     \caption{The sensitivity of the CMB temperature power spectrum  (top) and  matter power spectrum at $z=1$ (bottom) to variations in the two dGd parameters, $p_1$ and $p_2$.}
        \label{response}
   \end{center}
\end{figure}

Given the proposed shape for the  power spectrum (Eq.~\ref{pk}), we proceed by assessing the detectability of these fluctuations by CMB and large-scale data. 
  
\subsection{Results}

In this section, we study the imprints of dGd parameters on the {\it Planck}  measurement of the CMB power spectrum (Section~\ref{sec:cmb}) and make a forecast for the detectability of the dGd imprint with future large-scale data.
 Figure~\ref{response} compares the expected impact of the dGd parameters on the CMB (top)  and matter power spectrum (bottom) and illustrates where the maximum sensitivity of these observables to the parameters is.  
 It should be noted that $p_0$ is hardly distinguishable from the amplitude of primordial inflationary scalar perturbations $A_{\rm s}$ 
(assuming an almost scale-independent power spectrum). Therefore we do not consider it as a new parameter in our analysis.

\subsubsection{Cosmic Microwave Background}\label{sec:cmb}

\begin{figure}
   \begin{center}
     \includegraphics[scale=0.45]{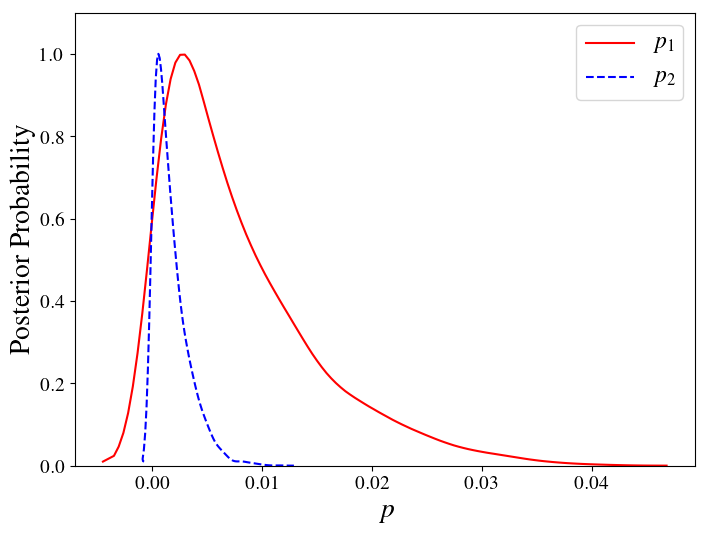}
     \caption{The posterior probability of the dGd parameters $p_1$ and $p_2$ using  {\it Planck} dataset. }
        \label{p1,p2}
   \end{center}
\end{figure}

We modify the publicly available code CosmoMC \footnote{https://cosmologist.info/cosmomc/} to take into account the contribution of the dGd induced inhomogeneities as a primordial source of inhomogeneities and leave the dGd parameters $p_1$ and $p_2$ as free parameters to be estimated by data. We assume uniform priors on these parameters and only require that the total dGd power spectrum, including contribution from both $p_1$ and $p_2$, be non-negative. Therefore, these two parameters are not separately restricted to non-negative values. As stated before, we use  {\it Planck} measurement of CMB temperature and polarization anisotropies \citep{Ade:2013xla}. 
We work in the $\Lambda$CDM theoretical framework, with the only modification possibly coming from the dGd phase transition.

\begin{table}
\centering
\caption{Best-fit parameters describing the initial conditions of the universe in the dGd model (both inflationary and dGd parameters) and their $1\sigma$ errors as measured by {\it Planck}.}
\label{tab:cmb}  
\begin{tabular}{lccr}
\hline
$p_1$ & $p_2$ & $n_{\rm s}$ & Log$(10^{10}A_{\rm s})$  \\
\noalign{\smallskip}\hline\noalign{\smallskip}
 $0.008^{+0.003}_{-0.008}$ & $0.002^{+0.001}_{-0.002}$& $0.9623^{+0.0075}_{-0.0053}$ & $3.0410^{+0.0161}_{-0.0239}$ \\
\noalign{\smallskip}\hline
\end{tabular}
\end{table}

   \begin{table*}
        \centering 
        \caption{Estimated errors on the dGd parameter pair forecasted for future observations of large-scale structure (Euclid, SKA1, and SKA2-like) surveys. In this analysis, the standard cosmological parameters are fixed.}
        \label{tab:lss1}
        \begin{tabular}{c|ccc|ccc|ccc}       
        \hline
          & \multicolumn{3}{c|}{Euclid-like}    & \multicolumn{3}{c|}{SKA1-like} & \multicolumn{3}{c}{SKA2-like}  \\
          & GC & WL & total & GC & WL & total &GC & WL & total\\ \hline
          $p_1$ & 0.0001 & 0.0008 & 0.0001 & 0.0015 &0.0021 & 0.0011 &0.0001&0.0005&0.0001\\ 
          $p_2$ & 0.0006 & 0.0016 & 0.0006 & 0.0062 &0.0027 & 0.0024 &0.0005&0.0009&0.0004\\
            \hline
        \end{tabular} 
    \end{table*}

  \begin{table}
        \centering
        \caption{Similar to Table~\ref{tab:lss1} but marginalized over the six standard cosmological parameters.}
        \label{tab:lss2}
        \begin{tabular}{c|ccc|ccc|ccc}
        \hline
          & \multicolumn{3}{c|}{Euclid-like}    & \multicolumn{3}{c|}{SKA1-like} & \multicolumn{3}{c}{SKA2-like}  \\
          & GC & WL & total & GC & WL & total &GC & WL & total\\ \hline
          $p_1$ & 0.002 & 0.014 & 0.002 & 0.024 &0.036 & 0.014 &0.002&0.008&0.001\\ 
          $p_2$ & 0.003 & 0.010 & 0.003 & 0.033 &0.022 & 0.012 &0.003&0.006&0.002\\ \hline
        \end{tabular}
    \end{table}

Our parameter set therefore includes the standard cosmological base parameters ($\Omega_bh^2$, $\Omega_ch^2$, $\theta$, $\tau$, $A_s$ and $n_s$, with $k_{\rm pivot}=0.05 {\rm Mpc}^{-1}$ as the pivot for scalar perturbations) along with the dGd parameters. We have assumed uniform priors on the dGd parameters, in the range  $[0,10]$. We find this prior range to be safe in the sense that it covers all the dGd parameter space with non-negligible likelihood, and the posterior is not cut in the edges of the parameter space due to prior biases. We take the number of relativistic species to be $N_\nu=3.046$ and assume the neutrinos to be massless. We have also assumed the primordial tensor perturbations have negligible contribution to CMB temperature and $E$-mode polarization anisotropies. The helium abundance is set from the BBN consistency relation. We use eight chains of parameters and use the Gelman and Rubin R-statistic to assess their convergence. We find that, with a total of about 32000 samples, the chains are converged with $R-1 < 0.01$. 

Table~\ref{tab:cmb} summarizes the results of this dGd-parameter measurement and Figure~\ref{p1,p2} shows the posterior probabilities of $p_1$ and $p_2$. 
The two dGd parameters are almost uncorrelated. That is expected since the two parameters affect different scales. Small scales, or large $k$'s, are most sensitive to $p_1$, while large-scales, or small $k$'s, are mostly affected by $p_2$  (see equation~\ref{pk} and Figure~\ref{response}).
The standard parameters also have little correlation with the dGd ones due to the distinct imprints they leave on the power spectrum and are therefore almost unchanged. 
The results indicate no deviation from the inflationary power-law spectrum in the form predicted by the dGd formalism. 

\subsubsection{large-scale Structure}

Features in the primordial power spectrum also leave imprints on matter distribution. We investigate the detectability of the dGd-induced features characterized by the two parameters $p_1$ and $p_2$ in future large-scale surveys.
In this work, we make forecast using simulations for the European Space Agency's Euclid mission, referred to as Euclid-like, and the Square Kilometer Array (SKA), with two different sets of proposed specifications, referred to as SKA1-like and SKA2-like. 
 In particular, we use the weak lensing (WL) and galaxy clustering (GC) probes, following specifications assumed in  
 \citep{Laureijs:2011gra, Amendola_2013,Santos:2015}.
We do a Fisher matrix analysis in the linear regime of perturbations assuming a near-Gaussian distribution for the parameter. The formalism and the details of the analysis are similar to the analysis thoroughly described in \citep{Esmaeilian:2021}.
 
Tables~\ref{tab:lss1} and \ref{tab:lss2} present the forecasted errors of the two dGd parameters for the various experimental scenarios used in this section, for the WL and GC probes, and with the standard cosmological parameters assumed fixed and free respectively.
The constraints from GC and WL are tightest from Euclid-like and SKA2-like and comparable to  {\it Planck} measurements (Table~\ref{tab:cmb}).  

\section{Summary and Discussion}

In this work, we investigated the observational consequences of a possible phase transition of the spacetime at the end of inflation, the so-called dGd phase transition. We simulated fluctuations in the inflaton field induced by this transition and found the fit to the corresponding power spectrum. 
The amplitudes of the various terms in the dGd power spectrum were considered as free parameters and were constrained by  {\it Planck} data. 
No significant deviations from the standard power-law inflationary power spectrum were found. 
The high-precision observations of the large-scale structures in the near future could improve these constraints. We made Fisher-based forecasts for Euclid and SKA-like surveys and found comparable bounds on the dGd parameters from the weak lensing and galaxy clustering probes. 

If deviations from pure inflationary power law are observed, the consistency of these perturbations with the dGd scenario could be tested by extracting the $\delta$s corresponding to each observed dGd parameter, $p_1$ and $p_2$, from Figure~\ref{pp}. The agreement of the deduced $\delta$'s (within the error bars) would imply the consistency of the observed deviation as seeded by an early dGd phase transition. 
The derived value for $\delta$ would also shed light on the physics of the phase transition through constraining  its duration $\tilde{t}$ (as discussed in Section~\ref{sec:res}), which itself depends on the free parameters of the theory  $\sigma$, $\lambda$ and $\Lambda$ through Equation~\ref{ttilde}.

\section{Acknowledgement}

Part of the numerical computations of this work was carried out on the computing cluster of the Canadian Institute for Theoretical Astrophysics (CITA), University of Toronto.

\nocite{*}
\bibliography{dgd}
\end{document}